\documentclass{emulateapj}

\usepackage{amssymb}
\usepackage{amsmath}
\usepackage{natbib}
\usepackage{txfonts}
\usepackage[colorlinks=true,citecolor=blue,linkcolor=blue]{hyperref}
\usepackage{microtype}
\usepackage{color}
\usepackage{graphicx}
\usepackage{epstopdf}

\bibpunct{(}{)}{;}{a}{}{,}

\def\mnras{MNRAS}
\def\apj{ApJ}
\def\apjl{ApJL}

\def\aap{A\&A}

\def\losalamos{1}
\def\icranet{2}
\def\roma{3}

\shorttitle{Induced gravitational collapse}
\shortauthors{Fryer, Rueda, Ruffini}

\begin{document}

\title{Hypercritical Accretion, Induced Gravitational Collapse, and Binary-Driven Hypernovae}

\author{Chris L.~Fryer\altaffilmark{\losalamos}, 
				Jorge A. Rueda\altaffilmark{\icranet,\roma}, 
				Remo Ruffini\altaffilmark{\icranet,\roma}}

\altaffiltext{\losalamos}{CCS-2, Los Alamos National Laboratory, Los Alamos, NM
87545}

\altaffiltext{\icranet}{ICRANet, 
                     P.zza della Repubblica 10, 
                     I--65122 Pescara, 
                     Italy}  
                     
\altaffiltext{\roma}{Dipartimento di Fisica and ICRA, 
                     Sapienza Universit\`a di Roma, 
                     P.le Aldo Moro 5, 
                     I--00185 Rome, 
                     Italy}

\begin{abstract}
The induced gravitational collapse (IGC) paradigm has been
successfully applied to the explanation of the concomitance of
gamma-ray bursts (GRBs) with supernovae (SNe) Ic. The progenitor is a
tight binary system composed by a carbon-oxygen (CO) core and a
neutron star (NS) companion.  The explosion of the SN leads to
\emph{hypercritical} accretion onto the NS companion which
reaches the critical mass, hence inducing its gravitational collapse
to a black hole (BH) with consequent emission of the GRB. The first
estimates of this process were based on a simplified model of the
binary parameters and the Bondi-Hoyle-Lyttleton accretion rate. We
present here the first full numerical simulations of the IGC
phenomenon. We simulate the core-collapse and SN explosion of CO stars
to obtain the density and ejection velocity of the SN ejecta. We
follow the hydrodynamic evolution of the accreting material falling
into the Bondi-Hoyle surface of the NS all the way up to its
incorporation to the NS surface. The simulations go up to
BH formation when the NS reaches the critical mass.
For appropriate binary parameters the IGC occurs in short timescales
$\sim 10^2-10^3$~s owing to the combined effective action of the
photon trapping and the neutrino cooling near the NS
surface. We also show that the IGC scenario leads to a natural
explanation for why GRBs are associated only to SN Ic with totally
absent or very little helium.
\end{abstract}

\keywords{Type Ic Supernovae --- Hypercritical Accretion --- Induced Gravitational Collapse --- Gamma Ray Bursts}

\maketitle

\section{Introduction}

Continued observations of massive stars have demonstrated that most,
if not all, massive stars are in binary systems
\citep[e.g.][and references therein]{
2004ApJ...610L.105S,2007ApJ...670..747K,
2012Sci...337..444S}.
A large fraction (50--75\%) of these systems are in tight binaries
that interact during the evolution
(e.g.~mass transfer, common envelope phase). The high binary fraction
has led to a growing consensus that most type Ib/Ic supernova
progenitors are produced in interacting binary
systems \citep{
1992ApJ...391..246P,
1998A&A...333..557D,2007PASP..119.1211F,2010ApJ...725..940Y}.
Since the type of SNe associated to long-duration GRBs are of type Ic
\citep{2011IJMPD..20.1745D}, it is not surprising 
that binaries, often involving interactions of a massive star with a 
compact companion, have been invoked to produce GRB-SNe to remove 
the hydrogen envelope, spin up the star, or both
\citep{1998ApJ...502L...9F,1999ApJ...526..152F,
2005ApJ...623..302F,2007Ap&SS.311..177V,2006ARA&A..44..507W,2007PASP..119.1211F}.

\begin{figure*}[!hbtp]
\centering
\includegraphics[width=0.8\hsize,height=8cm]{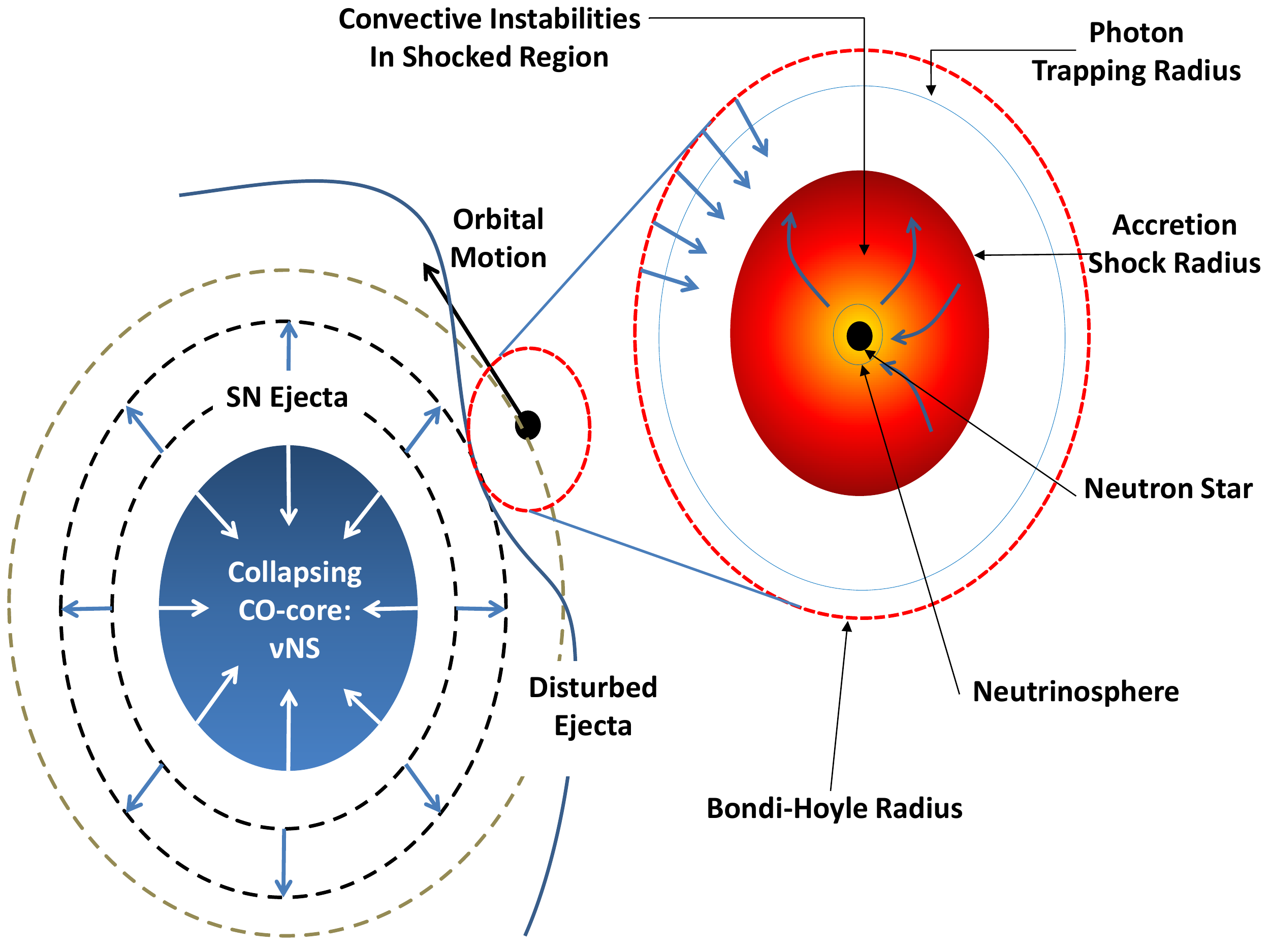}
\caption{Induced gravitational collapse scenario.}\label{fig:scenario}
\end{figure*}

The \emph{induced gravitational collapse} \citep[IGC,][]{2008mgm..conf..368R,2012ApJ...758L...7R} model requires a tight binary (produced in a common envelope phase) between a massive CO star (a star that has lost its hydrogen envelope and helium shell) and a NS companion. In this scenario, the SN explosion and the GRB occur following a precise time sequence (see Fig.~\ref{fig:scenario}): explosion of the CO core $\to$ hypercritical accretion onto the NS $\to$ the critical mass is reached $\to$ gravitational collapse to a BH is induced $\to$ emission of the GRB. The theoretical framework and the first estimates of the hypercritical accretion onto the NS as a function of the nature of the binary parameters were first presented in \citep{2012ApJ...758L...7R}. 

It has been clear since the analysis of GRB 090618 by \citet{2012A&A...543A..10I} that the entire emission of what has been traditionally called a GRB, instead of being a single event, is actually a \emph{multiepisodic} source whose understanding needs a time-resolved data scrutiny data. The IGC has been successfully applied to a class of energetic ($E_{\rm iso}\sim 10^{52}$-–$10^{54}$~erg) GRB-SNe. These systems, recently named binary-driven hypernovae \citep[BdHNe,][]{2014A&A...565L..10R}, evolve in a rapid sequence lasting a few hundreds of seconds in their rest-frame. Up to now, the IGC has been verified in a dozen of GRBs, all with cosmological redshift $z \leq 1$ \citep[see][and references therein]{2013A&A...552L...5P}, and very recently in one of the farthest observed sources, GRB 090423 at $z = 8.2$ \citep{2014arXiv1404.1840R}. These systems are characterized by four distinct episodes, each with specific signatures in its spectrum and luminosity evolution.

\textit{Episode 1}: first part of the emission, it presents a soft X-ray spectrum with peak energies $<100$~keV and is generally time-separated from the rest of the emission. It shows a complex spectrum which at times presents a thermal component. Physically, it has the imprint of the onset of the SN in a tight binary system with the companion NS. Its emission mainly originates from the hypercritical accretion, $\dot{M}\sim 10^{-2}~M_\odot$~s$^{-1}$, of the SN ejecta onto the NS. 

\textit{Episode 2}: second part of emission observed with peak energies $\sim 100$~keV--$1$~MeV. It is the canonical GRB emission originated from the gravitational collapse of the NS to a BH. The dynamics of the evolution of the highly relativistic (Lorentz factor $\Gamma \gtrsim 10^2$) $e^+e-$ plasma, which engulfs baryonic matter and interacts with the circumburst medium (CBM), follows the fireshell model which takes into account the special relativistic effects and the plasma rate equation \citep[see][and references therein]{2011IJMPD..20.1797R}.

\textit{Episode 3}: the previously called \emph{afterglow emission}, visible in optical, X-rays, and with a high energy component up to GeV energies, which observationally starts at the end of the GRB prompt emission. Independently from the features of the Episode 2 and its energetics, the Episode 3 appears to have a most remarkable scaling law and a universal behavior for all the canonical GRBs. In the Swift-XRT lightcurve it consists, starting at the end of the GRB prompt, in a steep decay followed by a plateau and a late power-law decay \citep{2013A&A...552L...5P}. The late X-ray luminosities of BdHNe, in their rest-frame energy band $0.3$--$10$~keV, evidence a common power-law behavior, $L_X\propto t^\alpha$, with a constant decay index clustered around $\alpha = -1.5 \pm 0.2$. Such a constant afterglow decay represents an authentic nested structure \citep{2014A&A...565L..10R} in the late X-ray emission of GRB-SNe and it has been indicated as the qualifying feature for a GRB to be a BdHNe family member. The identification of GRB 090423 at $z=8.2$ as a BdHN \citep{2014arXiv1404.1840R} implies that SN events, leading to NS formation, can occur already at 650~Myr after the Big Bang. The above opens the way to consider the late X-ray power-law as a possible distance indicator.

\textit{Episode 4}: emergence of the SN emission after $\sim 10$--$15$ days from the occurrence of the GRB, in the source rest-frame. It has been observed for almost all the sources fulfilling the IGC paradigm with $z \sim 1$, for which current optical instrumentation allows their identification.

The first estimates of the IGC process
\citep{2012ApJ...758L...7R,2012A&A...543A..10I,2012A&A...538A..58P,2013A&A...551A.133P,2013A&A...552L...5P,2013arXiv1311.7432R}
were based on a simplified model of the binary parameters and the
Bondi-Hoyle-Lyttleton accretion framework. The aim of this Letter is
to better constrain the binary characteristics that lead to the IGC
phenomenon (Episode 1) using more detailed supernova explosions coupled with
models based on simulations of hypercritical accretion in supernova
fallback \citep{1996ApJ...460..801F,2009ApJ...699..409F}.
We consider numerical simulations of collapsing CO cores leading to SN
Ic in order to calculate realistic profiles for the density
and ejection velocity of the SN outer layers. We
follow the hydrodynamic evolution of the accreting material falling
into the Bondi-Hoyle accretion region all the way up to its
incorporation onto the NS surface. 
\section{Binary progenitor}

The hypercritical accretion onto the NS from
the SN ejecta in the IGC scenario can be estimated using the Bondi-Hoyle-Lyttleton
formalism
\citep{1939PCPS...35..405H,1944MNRAS.104..273B,1952MNRAS.112..195B}:
\begin{equation}\label{eq:Mdot}
\dot{M}_{\rm BHL}=4 \pi r_{\rm BHL}^2 \rho (v^2 +c_s^2)^{1/2}\, ,
\end{equation}
where $\rho$ is the density of the SN ejecta, $v$ is the ejecta
velocity in the rest frame of the NS (this includes a
component from the ejecta velocity, $v_{\rm ej}$ and another component
from the orbital velocity of the NS, $v_{\rm orb}$), $c_s$
is the sound speed of the SN ejecta, and $r_{\rm BHL}$ is the Bondi
radius:
\begin{equation}
r_{\rm BHL}=\frac{G M_{\rm NS}}{v^2 +c_s^2}
\end{equation}
where $G$ is the gravitational constant and $M_{\rm NS}$ is the
NS mass. Both the velocity components, $v_{\rm orb}, v_{\rm
  ej}$, are typically much higher than the sound speed.  The ejecta
velocity as a function of time is determined by the explosion energy
and the nature of the SN explosion.  The orbital velocity depends upon
the orbital separation, which in turn depends upon the radius of the
CO star and the binary interactions creating the tight-orbit binary
just prior to the explosion of the CO core. The effect of the NS magnetic field can be neglected for $\dot{M}>2.6\times 10^{-8}~M_\odot$~s$^{-1}=0.8~M_\odot$~yr$^{-1}$ \citep{1996ApJ...460..801F,2012ApJ...758L...7R}



\begin{figure}[!hbtp]
\centering 
\plotone{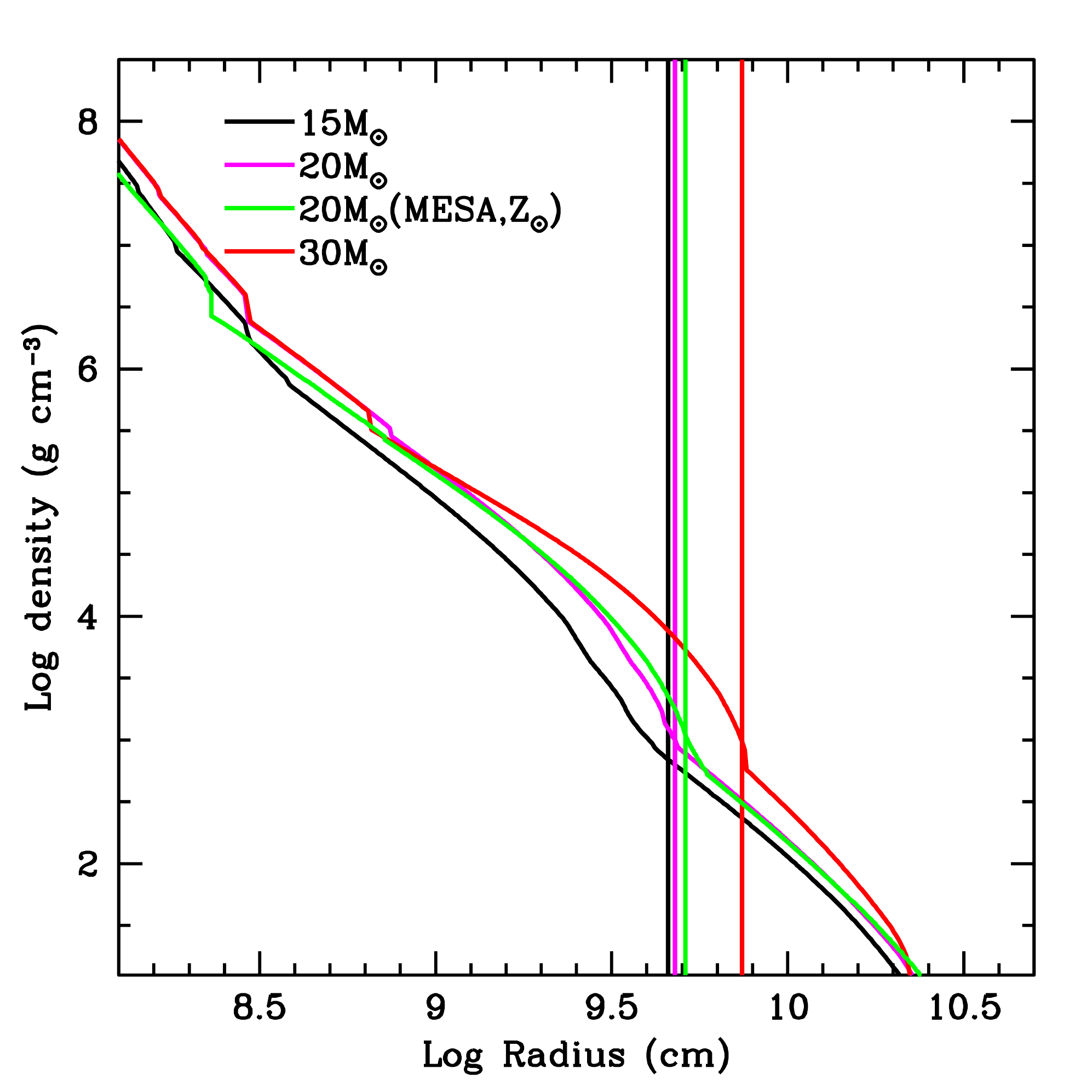}
\caption{Density profile of different CO core progenitors with $M_{\rm
    ZAMS}=$15, 20, and 30~$M_\odot$ for low-metallicity
  (Z=0.0001~Z$_\odot$) stars using the Kepler stellar evolution
  code \citep{2002RvMP...74.1015W}. For comparison, we include the
  density structure of a solar metallicity star produced by the MESA
  code (Sam Bond, in preparation).  The vertical lines show the radius
  of the CO core.  As we shall see below, the accretion rate is
  extremely sensitive to the structure of the star.}
\label{fig:progenitor}
\end{figure}

The density evolution of the SN ejecta near the NS companion
depends upon the SN explosion and the structure of the progenitor
immediately prior to collapse.  In Fig.~\ref{fig:progenitor}, we show
the density profile for three different low-metallicity stars with
initial zero-age main sequence masses of $M_{\rm ZAMS}=$15, 20, and
30~$M_\odot$ \citep{2002RvMP...74.1015W}. We designate the edge of the CO core in all of
these stars. The density profile depends upon both on the initial
conditions of the star (metallicity, initial mass, rotation) as well
as the the stellar evolution code used (in this case, KEPLER).  The
density profile of a 20~$M_\odot$, solar metallicity star (Sam Jones,
in preparation), is obtained using the MESA code. The IGC model assumes
that both the hydrogen and helium layers are removed prior to collapse. There is a 3--4 order of magnitude pressure jump between
the CO core and helium layer, indicating that the star will not expand significantly 
when the helium layer is removed.  Comparisons of KEPLER models with the stripped CO 
cores from \cite{2010ApJ...719.1445M} suggest that, for some stellar evolution codes, 
the CO cores could be 1.5-2 times larger.  We will discuss this effect on the accretion 
rate below.

The compactness of the CO core (see Fig.~\ref{fig:progenitor}) is such that there is no Roche lobe overflow\footnote{The Roche lobe radius is \citep{1983ApJ...268..368E}: $R_{\rm L,CO}\approx 0.49 q^{2/3}/[0.6 q^{2/3} + \ln(1 + q^{1/3})]$, where $q=M_{\rm CO}/M_{\rm NS}$.} prior to the SN explosion. For instance, for a CO core progenitor with $M_{\rm ZAMS}=15~M_\odot$ ($M_{\rm CO}\approx 5~M_\odot$, $R_{\rm CO}\approx 3\times 10^9$~cm) no Roche lobe overflow occurs for binary periods $P\geq 2$~min, or binary separation $a\geq 6\times 10^9$~cm, for a NS companion mass $M_{\rm NS}\geq 1.4~M_\odot$.


\section{Binary driven hypercritical accretion}

\begin{figure}[!hbtp]
\centering
\plotone{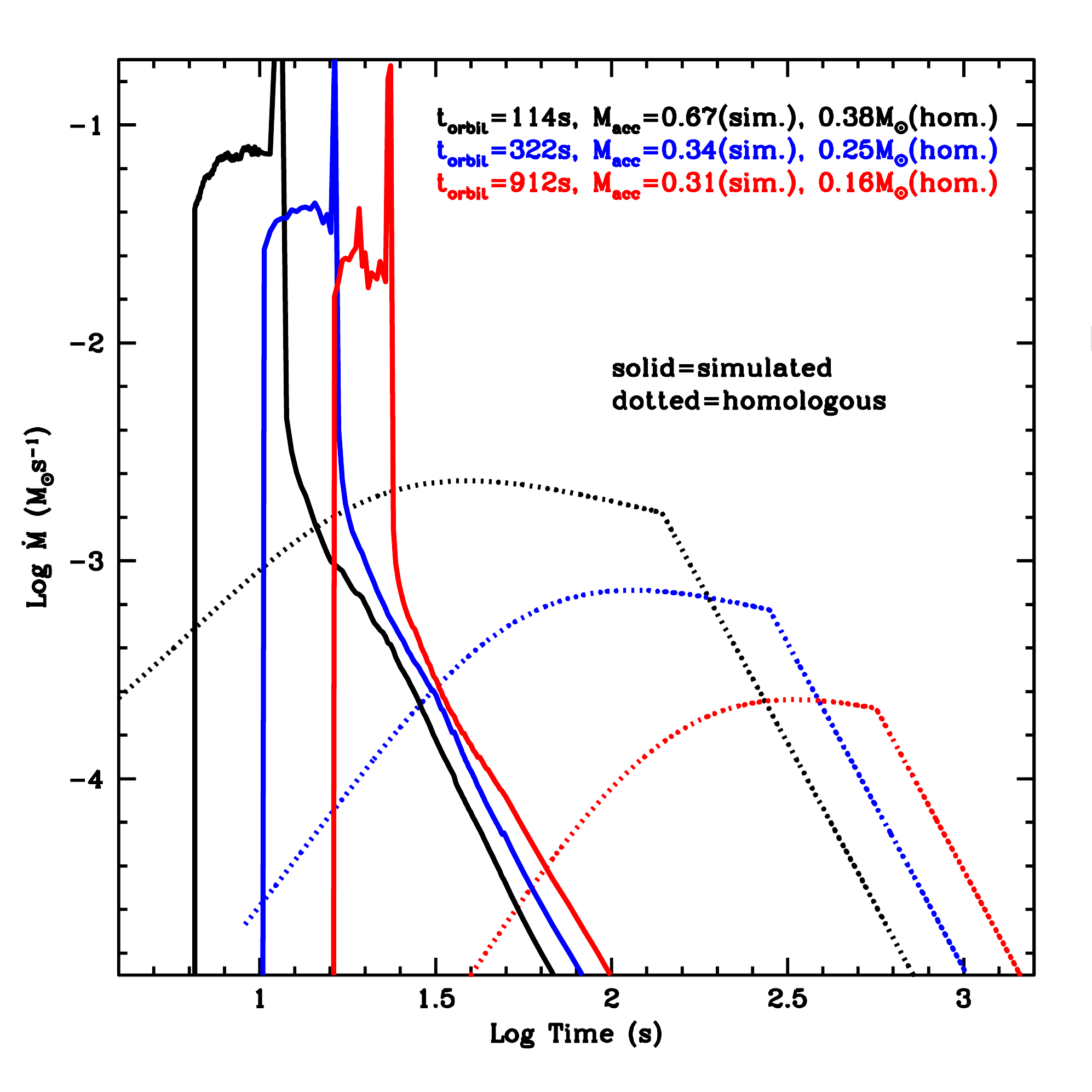}
\caption{Hypercritical accretion for selected separation distances and for a star progenitor of $20~M_\odot$ using 
our two approaches for the explosive engine.  The supernova shock increases the density of the outgoing material, 
producing a pile up at the shock that leads to a spike in the accretion rate over a brief (few second) period, a 
much sharper accretion profile than our $\gtrsim 100\,$s accretion time for our homologous outflow models.}
\label{fig:BHaccretion}
\end{figure}

To derive the hypercritical accretion onto the NS, we must
implement an explosion model.  Here we take two approaches.  The first
is to assume a homologous outflow with a set explosion energy
on the progenitor star structure.  For comparison, we also use a
second approach that follows the collapse, bounce, and explosion of
the 20$M_\odot$ progenitor discussed above using the parameterized
model developed to study a range of SN explosion energies for fallback
and SN light-curves \citep{2013ApJ...773L...7F}.  The calculation uses
a 1D core-collapse code \citep{1999ApJ...516..892F} to follow the
collapse and bounce and then injects energy just above the
proto-NS to drive different SN explosions mimicking the
convective-engine paradigm. With this progenitor and explosion, we
produce an example density and velocity evolution history at the
position of the Bondi-Hoyle surface of our binary
companion. Fig.~\ref{fig:BHaccretion} shows the Bondi-Hoyle infall
rate from both our homologous outflow and our simulated SN models for
a range of orbital separations (the innermost separation is determined
to be just high enough so that the CO star does not overfill its Roche
lobe).  In our simulated explosion, the density piles up in the shock,
producing a much sharper burst of accretion onto the NS.
The accretion rate can be an order of magnitude higher in these models, but for a much shorter time such that the
total mass accreted is only less than 2 times higher.  

This infall rate is well above the Eddington rate and will be reset to this rate if the assumptions of the Eddington accretion limit apply.  The Eddington rate is derived assuming that the energy released when material accretes onto a compact object is released in photons and these photons exert pressure on the infalling material, reducing the accretion rate.  The Eddington accretion limit, or critical accretion rate, makes a series of assumptions: the potential energy is released in the form of photons, the inflowing material and outflowing radiation is spherically symmetric, the photons are not trapped in the flow and can deposit momentum to the inflowing material, and the opacity is dominated by electron scattering.  For a wide variety of accreting X-ray binaries, the Eddington limit seems to hold (at the order of magnitude level). But many of these assumptions break down for accretion rates as high as the ones achieved in the IGC scenario, hence hypercritical.

First and foremost, the photons in the hypercritical IGC accretion rates are almost certainly trapped in the flow. \citet{1989ApJ...346..847C} derived the trapping radius where photons emitted diffuse outward at a slower velocity than infalling material flows inward:
\begin{equation}
r_{\rm trapping} = min [(\dot{M}_{\rm BHL} \kappa)/(4 \pi c),r_{\rm BHL}]
\end{equation}
where $\kappa$ is the opacity (in cm$^2$~g$^{-1}$) and $c$ is the speed of light. If the trapping radius is near or equal to the Bondi-Hoyle radius, the photons are trapped in the flow and the Eddington limit does not apply.  This hypercritical accretion has been studied in detail for common envelope scenarios where $\kappa$ is likely to be dominated by electron scattering.  However, in SN fallback \citep{1999ApJ...511..885F} and the IGC model, heavy elements are not completely ionized and lines can significantly increase the opacity.  Following \citet{2013HEDP....9..369C}, we estimate for our CO core a Rosseland mean opacity roughly $5\times 10^3$~cm$^2$~g$^{-1}$, a factor $\sim 10^4$ higher than electron scattering. This means that the trapping radius is higher for the IGC model. Combined with our high accretion rates, it is clear that the Eddington limit does not apply in this scenario and hypercritical, largely Super-Eddington accretion, occurs. The inflowing material shocks as it piles up onto the NS, producing an atmosphere on top of the NS \citep[for details, see][]{1972SvA....16..209Z,1989ApJ...346..847C,1991ApJ...376..234H,1996ApJ...460..801F}. As the atmosphere compresses, it becomes sufficiently hot to emit neutrinos which cool the infalling material, allowing it to be incorporated into the NS. For details of the simulation of this process, we refer the reader to \citep{1973PhRvL..31.1362R,1996ApJ...460..801F,2009ApJ...699..409F}.

\begin{deluxetable}{lcccc}[!hbtp]
\centering
\tabletypesize{\scriptsize}
\tablecaption{Hypercritical Accretion Mass in the IGC scenario \label{tab:data}}
\tablewidth{0pt}
\tablehead{\colhead{Progenitor} & \multicolumn{4}{c}{$M_{\rm acc}^a$~$(M_\odot)$, $t_{\rm acc}$~(s)} \\
\colhead{ZAMS Mass} & \colhead{$a_{\rm orbit}/a_{\rm min}^{b}=1$} & \colhead{2} & \colhead{4} & \colhead{8}}
\startdata
15~$M_{\odot}$ & 0.24,~160 & 0.15,~400 & 0.085,~600 & 0.042,~1300\\
20~$M_{\odot}$ & 0.38,~ 150 & 0.25,~250 & 0.16,~600 & 0.096,~1200\\
20~$M_{\odot}$$^{c}$ & 0.67,~5 & 0.34,~6 & 0.31,~7 & 0.17,~7\\
20~$M_{\odot}$$^{d}$ & 0.084,~150 & 0.058,~250 & 0.032,~600 & 0.001,~1200\\
30~$M_{\odot}$ & 0.62,~800 & 0.42,~2000 & 0.28,~3700 & 0.16,~8000
\enddata
\tablenotetext{a}{Total accretion at super-Eddington rates in $M_\odot$.}
\tablenotetext{b}{$a_{\rm min}$: minimum orbital separation such that the CO core does not fill its Roche lobe.}
\tablenotetext{c}{Simulated with the KEPLER code.}
\tablenotetext{d}{Solar metallicity star, simulated with the MESA code.}
\end{deluxetable}

Table~\ref{tab:data} shows the total mass accreted ($M_{\rm acc}$) for selected
orbital separations and progenitor masses using different stellar
evolution codes and different models (homologous vs. simulated) of the
SN explosion. We also indicate the time interval ($t_{\rm acc}$) in which the accretion rate is integrated to obtain $M_{\rm acc}$. 
For these systems, the accretion rate is largely hypercritical exceeding $10^{-3}~M_\odot$~s$^{-1}$, so we expect a
fraction of these systems to push beyond the maximum NS mass
and collapse to a BH. Note that for the helium star systems, the accretion rate is not high enough to produce an IGC.  
If the radius of the CO core was twice that of our models (see discussion on the \cite{2010ApJ...719.1445M} models), 
our peak accretion rates would correspond to the $a_{\rm orbit}/a_{\rm min}^{b}=2$ values.

As material piles onto the NS and the atmosphere radius, the accretion shock moves outward.  The accretion shock weakens as it moves out and the entropy jump (derived from the shock jump conditions) becomes smaller.  This creates an atmosphere that is unstable to Rayleigh-Taylor convection.  Simulations of these accretion atmospheres show that these instabilities can accelerate above the escape velocity driving outflows from the accreting NS with final velocities approaching the speed of light, ejecting up to 25\% of the accreting material \citep{2006ApJ...646L.131F,2009ApJ...699..409F}. The entropy of the material at the base of our atmosphere, $S_{\rm bubble}$, is given by \citep{1996ApJ...460..801F}:
\begin{equation}
S_{\rm bubble}=38.7~\left(\frac{M_{\rm NS}}{2M_\odot} \right)^{7/8} \left(\frac{\dot{M}_{\rm
  BHL}}{0.1~M_\odot {\rm s}^{-1}}\right)^{-1/4}\left(\frac{r_{\rm NS}}{10^6~{\rm cm}}\right)^{-3/8}
\end{equation}
$k_B$ per nucleon, where $r_{\rm NS}$ is the radius of the NS.  The corresponding temperature of the bubble, $T_{\rm bubble}$, is:
\begin{equation}
T_{\rm bubble} = 195~S_{\rm bubble}^{-1} \left(\frac{r_{\rm NS}}{10^6~{\rm cm}}\right)^{-1}.
\end{equation}
For the typical hypercritical accretion conditions of the ICG, the temperature of the bubble when it begins to rise is $T_{\rm bubble}\sim$5~MeV. If it rises adiabatically, expanding in all dimensions, its temperature drops to 5~keV at a radius of $10^9$~cm, far too cool to observe.  However, if it is ejected in a jet, as simulated by \citet{2009ApJ...699..409F}, it may expand in the lateral direction but not in the radial direction, so $\rho \propto r^{2}$ and $T \propto r^{-2/3}$. In this scenario, the bubble outflow would have $T_{\rm bubble}\sim 50$~keV at $10^9$~cm and $T_{\rm bubble}\sim 15$~keV at $6\times10^9$~cm. This could explain the temperature and size evolution of the blackbody emitter observed in the Episode 1 of several BdHNe \citep[see, e.g.,]{2012A&A...543A..10I,2012A&A...538A..58P,2013A&A...551A.133P,2013A&A...552L...5P,2013arXiv1311.7432R}. For instance, the blackbody observed in Episode 1 of GRB 090618 \citep{2012A&A...543A..10I} evolves as $T\propto r^{-m}$ with $m=0.75 \pm 0.09$, whose lower value is in striking agreement with the above simplified theoretical estimate. We are currently deepening our analysis of the possible explanation of the thermal emission observed in Episode 1 of BdHNe as due to the convective instabilities in the accretion process. However, this is out of the scope of this work and the results will be presented elsewhere.

\section{Discussion}

While in this Letter we address simulations of the Episode 1 of the IGC, let us shortly outline some recent progress in the understanding the structure of Episode 3 which may become complementary to this work: a) the remarkable scaling laws in the X-ray luminosity in all BdHNe \citep[see][for details]{2013A&A...552L...5P,2014A&A...565L..10R}; b) the very high energy emission, all the way up to 100~GeV in GRB 130427A, as well as the optical one, follow a power-law behavior similar to the X-ray emission described above. The corresponding spectral energy distribution is also described by a power-law function with quite similar decay indexes \citep{2014arXiv1405.5723R}. These results clearly require a common origin for this emission process; c) an X-ray thermal component has been observed in the early phases of the Episode 3 of GRBs 060202, 060218, 060418, 060729, 061007, 061121, 081007, 090424,100316D, 100418A, 100621A, 101219B and 120422A \citep{2011MNRAS.416.2078P,2012MNRAS.427.2950S,2013ApJ...771...15F}. This feature has been clearly observed in GRB 090618 and GRB 130427A, implying a size of the emission region of $10^{12}$--$10^{13}$~cm expanding at velocity $0.1<v/c<0.9$, hence a bulk $\Gamma$ Lorentz factor $\lesssim 2$ \citep{2014A&A...565L..10R,2014arXiv1405.5723R}. 

Recently, \citet{2014A&A...565L..10R} raised the
possibility of using the nuclear decay of ultra-heavy r-process
nuclei, originated in the close binary phase of Episode 1, as an
energy source of the Episode 3. These processes lead to a power-law
emission \citep[see, e.g.,][]{2013ApJ...774...25K} with decay index
similar to the one observed in Episode 3. The total energy emitted in
the nuclear decays is also in agreement with the observations in the
Episode 3 of BdHNe. r-process avalanches in BdHNe could also originate
from a similar mechanism as the one outlined by
\citet{2006ApJ...646L.131F} in SN fallback. An additional possibility
to generate the scale-invariant power-law in the luminosity evolution
and spectrum are the type-I and type-II Fermi acceleration processes
\citep{1949PhRv...75.1169F} during the evolution of the SN
remnant. The application of the Fermi acceleration mechanisms has two
clear advantages; the generation of the aforementioned power-law
behaviors, and to solve the longstanding problem formulated by Fermi
of identifying the injection source to have his acceleration mechanism
at work at astrophysical scales.

We have advanced our estimates of the NS accretion rate
within the IGC model, which leads to BdHNe with all the above features.  Our estimates assume that the
Bondi-Hoyle-Lyttleton formalism is valid for our calculations. Although it has been shown that this formalism is valid in
steady-state systems \citep[see][and references therein]{2004NewAR..48..843E}, the IGC model, with its
time-variable conditions may push the validity of these assumptions.
Full accretion models are required in order to validate our results and/or to produce more reliable accretion rates.

It appears from observations that a necessary condition to produce a
GRB-SN is that the pre-SN core is fully absent of or has very little
helium. We have shown that the IGC process provides a natural
explanation for that condition: hypercritical accretion rates are
favored by the presence of a compact CO core since it leads to tighter
binaries and produces higher opacities of the ejecta which favors the
photon trapping. We showed that helium cores do not trigger enough
hypercritical accretion onto the NS companion to produce an IGC.  A
number of mechanisms have been proposed to remove this material (a
common problem for most GRB scenarios): stellar winds (the difficulty
with this model is removing just the helium layer and not a
considerable portion of the CO core), mass transfer (only low-mass
helium cores undergo a helium giant phase, so conditions for mass
transfer or common envelope phases may be difficult to reproduce),
enhanced mixing allow fusion to consume the helium layer
\citep{2013ApJ...773L...7F}. Detailed simulations of the binary
evolution up to the formation of binary systems conforming with the
IGC conditions are needed in order to assess this fundamental
question.

\end{document}